\documentclass[aps, prl, reprint, groupedaddress, superscriptaddress, amsmath]{revtex4-1}
\usepackage{graphicx, color}
\usepackage{amsmath}
\usepackage{epstopdf}

\newcommand{\intit}{\int_{-\infty}^{t}}

\begin{document}
\title{Field-cycle-resolved photoionization in solids}
\date{\today}

\author{P. A. Zhokhov}
\affiliation{Department of Physics and Astronomy, Texas A\&M University, 77843 College Station TX, USA}
\affiliation{Russian Quantum Center, 143025 Skolkovo, Moscow Region, Russia}

\author{A. M. Zheltikov}
\affiliation{Department of Physics and Astronomy, Texas A\&M University, 77843 College Station TX, USA}
\affiliation{Russian Quantum Center, 143025 Skolkovo, Moscow Region, Russia}
\affiliation{Physics Department, International Laser Center, M.V. Lomonosov Moscow State University, 119992 Moscow, Russia}

\begin{abstract}
The Keldysh theory of photoionization in a solid dielectric is generalized to the case of arbitrarily short driving pulses of arbitrary pulse shape. We derive a closed-form solution for the nonadiabatic ionization rate in a transparent solid with a periodic dispersion relation, which reveals ultrafast ionization dynamics within the field cycle and recovers the key results of the Keldysh theory in the appropriate limiting regimes.
\end{abstract}
\maketitle

In his seminal 1964 paper \cite{Keldysh1965}, Keldysh has presented his celebrated formulas for photoionization, providing a uniform description of multiphoton and tunneling ionization. Over the next five decades, the Keldysh theory of photoionization has been pivotal to the research in laser science, providing a commonly accepted framework for a quantitative analysis of ionization in a remarkable diversity of light--matter interaction phenomena, including laser-induced breakdown \cite{Bloembergen1974, Lenzner1998}, high-order harmonic \cite{Brabec2000} and terahertz \cite{Tonouchi2002} generation, as well as filamentation of ultrashort light pulses \cite{Couairon2007, Berge2007}.
While the original Keldysh formulas were intended to describe photoionization in a continuous-wave field, several elegant approaches have been proposed \cite{Perelomov1966, Ammosov1986, Yudin2001} in the context of rapidly progressing ultrafast technologies \cite{Goulielmakis2007} and attosecond science \cite{Corkum2007}, to include the wave-packet nature of ultrashort driver pulses inducing an ultrafast ionization of gases.
These approaches help identify new field-cycle-sensitive phenomena in electron tunneling \cite{Uiberacker2007, Balciunas2013} and develop novel experimental methods for all-optical detection of electron tunneling dynamics \cite{Verhoef2010,Mitrofanov2011}.

Extension of the Keldysh model to ultrafast photoionization in solids is a standalone challenge in quantum physics. Meeting this challenge not only requires an adequate treatment of broadband driver fields, but also calls for a revision of the standard, hyperbolic model of the electron band structure adopted in the Keldysh formalism. The hyperbolic band model enables an accurate description of weak-field optical properties of solids \cite{Kane1959, Kane1961}, but fails in the strong-field regime, 
where effects of zone edges become significant. A Schr\"odinger-equation treatment with a 1D cosine-type dispersion \cite{Bonch-Bruevich1982,Hawkins2013} has been shown to partially address this problem, offering an adequate framework for the numerical analysis of an important class of ultrafast ionization effects in solids \cite{Schiffrin2013}. Still, in the lack of a closed-form solution for the photoionzation rate valid for ultrashort pulses of arbitrary shape, the physical intuition based on the Keldysh theory of photoionization of solids often has to be pushed beyond the range where this theory is rigorously valid, for the sake of compact semianalytical description and overall physical clarity \cite{Mitrofanov2011, Serebryannikov2009}. 

Here, we derive a closed-form solution for the nonadiabatic ionization rate in a transparent solid, which can be used not only to calculate the probability of ionization in the wake of the pulse and after each field cycle, but also to analyze the behavior of the ionization rate within the field cycle. Our analysis presented below in this paper reveals ultrafast ionization dynamics within the field cycle and recovers the results of the Keldysh theory within its range of applicability.


Our treatment is based on a two-band approximation of the electron band structure. The electron wave functions in the conduction and valence bands are written, following Keldysh \cite{Keldysh1965}, in the form of Volkov-type \cite{Volkov1935} wave functions:
\begin{equation} \label{eq:psi}
\psi_{c,v}\left(\vec{p},\vec{r},t\right) = u_{c,v}\left(\vec{{p}'}(t), \vec{r} \right) e^{i \vec{{p}'}\cdot\vec{r} - i\int_{-\infty}^{t}\mathcal{E}_{c,v}\left(\vec{{p}'}(\tau)\right)d\tau+i\varphi_{c,v}},
\end{equation}
where $\vec{{p}'}(t) = \vec{p} + \vec{A}(t)$,$\ u_{c,v}\left(\vec{p}, \vec{r}\right)$ are the Bloch wave functions of the conduction $(c)$ and valence $(v)$ bands, $\vec{r}$ is the position vector, $\vec{p}$ is the crystal quasi-momentum, $\vec{A}(t) = -\int_{-\infty}^{t} \vec{{E}}(t_1)dt_1$ is the vector potential, $\vec{E}(t)$ is the linearly polarized electric field with polarization direction $\vec{e}$, and $\mathcal{E}_{c,v}\left(\vec{p}\right)$ are the energies of conduction ($c$) and valence ($v$) bands. Here, unlike the Keldysh theory, the driving field is not assumed to be monochromatic and can have an arbitrary waveform. We also include random fluctuations of the phase $\varphi_{c,v}\equiv\varphi_{c,v}(t)$ to account for decoherence processes \cite{Kuehn2010}.
Assuming that the valence band is fully occupied and the conduction band is empty before the driving field is switched on, we write the probability amplitude for the electron transition to the conduction band (CB) as 

\begin{equation} \label{eq:L}
L\left(\vec{p},t\right) = \mathcal{N}\int_{-\infty}^{t} V_{cv}(\vec{p'}(t'))E(t') e^{-i\int_{-\infty}^{t'}\mathcal{E}\left(\vec{{p}'}(\tau)\right)d\tau +i\varphi(t')},
\end{equation}
where $\varphi(t') \equiv \varphi_c(t') - \varphi_v(t')$, $\mathcal{E}({\vec{p}}) \equiv \mathcal{E}_c({\vec{p}}) - \mathcal{E}_v({\vec{p}})$,  $V_{cv}\left(\vec{p}\right) \equiv \int u_{c,\vec{p}}^*(\vec{r}) \left(\vec{r}\cdot\vec{e}\right) u_{v,\vec{p}}(\vec{r}) d^3 \vec{r}$, and $\mathcal{N}$ is the normalization factor. 

The population of the conduction band is found as
\begin{equation} \label{eq:W}
W_c (t) = \left\langle \int_{BZ} d^D \vec{p} |L(\vec{p}, t)|^2 \right\rangle,
\end{equation}
where the integration is over the first Brillouin zone (BZ) of $D$-dimensional solid and $\langle ..\rangle$ denotes ensemble average.

Up to this point, we have closely followed the derivation by Keldysh \cite{Keldysh1965}.
The next step, however, will substantially deviate from the Keldysh treatment. In the Keldysh theory, integration in time in Eq. \eqref{eq:L} for a monochromatic laser field is followed by the integration in $p$ in Eq. \eqref{eq:W}. The approach that we adopt in this work is different, as we integrate over the momentum in Eqs. \eqref{eq:L} and \eqref{eq:W} first, making no assumption concerning the waveform of the driving field. This change in the order of integration in Eqs. \eqref{eq:L} and \eqref{eq:W} is central for our analysis, as it helps calculate the CB population for a driver pulse of general form.

To perform integration in $\vec{p}$ in Eq. \eqref{eq:W}, we need to specify the explicit form of dispersion $\mathcal{E}\left(\vec{p}\right)$. The Kane-type dispersion used in the Keldysh treatment is known to provide an adequate approximation for the dispersion around the zone center, but fails to describe periodicity of dispersion in the momentum space and  dispersion bending near the zone edges. This leads to serious difficulties for high field intensities, when the effects of zone edges may become significant \cite{Hawkins2013,Ghimire2010}. Here, we address these issues by using a cosine-type dispersion \cite{Bonch-Bruevich1982}
\begin{equation} \label{eq:cosine_bandD}
\mathcal{E}\left(\vec{p}\right) = \mathcal{E}_g + \Delta - \frac{\Delta}{D}\sum_{k=1}^D\cos(d_k p_k).
\end{equation}
Here $\mathcal{E}_g$ is the band gap, $p_k$ is the projection of the momentum on the $k$th Cartesian coordinate axis, $d_k$ are the lattice constants,
$\Delta = D/(m_k d_k)$, with $m_k$ being the effective electron--hole mass along the $k$th principal axis of the effective mass tensor.

Introducing $\vec{\mu} = \left\lbrace A_k d_k \right\rbrace_{k=1}^D$ and $\vec{x} = \left\lbrace p_k d_k \right\rbrace_{k=1}^D$, where $A_k$ is the $k$th Cartesian component of the vector potential,  we can represent the CB population at time $t$ as
\begin{equation} \label{eq:W_general}
\begin{split}
&W_c(t) = |\mathcal{N}|^2\intit dt_1 \intit dt_2 \int_{BZ} d^D \vec{x} \times \\
& E(t_1) E(t_2) V_{cv}\left(\vec{x}+\vec{\mu}(t_1)\right)V_{cv}^*\left(\vec{x}+\vec{\mu}(t_2)\right)  \times  \\
&e^{-i\int_{t_1}^{t_2} \mathcal{E}\left(\vec{x}+\vec{\mu}(\tau)\right) d\tau} \left\langle e^{i \varphi(t_2)-i\varphi(t_1)} \right\rangle.
\end{split}
\end{equation}
The integrals in $t_1, t_2$ in Eq. \eqref{eq:W_general} are dominated by the contributions from the saddle points of the oscillating exponent, corresponding to the pole where $V_{cv}(\vec{x} + \vec{\mu}(t))$ has a residue that is independent of the specific form of $V$ as a function of $\vec{x}$. Therefore, assuming that random fluctuations of the phase $\varphi$ are stationary and independent of $\vec{x}$, we can perform integration in $d^D x$ in Eq. \eqref{eq:W_general} to find
\begin{equation} \label{eq:Wfinal}
W_c(t) = |\mathcal{\tilde{N}}|^2\intit dt_1 \intit dt_2 E(t_1) E(t_2) G(t_1, t_2)
\end{equation}
where
\begin{equation}
\begin{split}
 G(t_1,t_2) =  e^{-i(\mathcal{E}_g + \Delta)(t_2 - t_1)} B\left(|t_2-t_1|\right)\times \\
 \prod_{k=1}^D \int_{-\pi}^{\pi} dx_k \exp\left\lbrace{ i \mathrm{Re}\Phi_k \cos x_k - i \mathrm{Im}\Phi_k \sin x_k}\right\rbrace,
 \end{split}
\end{equation}
\begin{equation} 
B(|t_2 - t_1|) = \left\langle e^{i \varphi(t_2)-i\varphi(t_1)} \right\rangle, \label{eq:B}
\end{equation}
$\Phi_k = \frac{\Delta}{D}\int_{t_1}^{t_2} \exp \left\lbrace i \mu_k(\tau)\right\rbrace d\tau$ and $\mathcal{\tilde{N}}$ is the field-independent normalization factor. The factor $B(\tau)$ includes decoherence, with $B(0) = 1$ and $B(\tau \rightarrow \infty) \rightarrow 0$.
Then, using $J_0(z) = \frac{1}{2\pi}\int_{-\pi}^{\pi} e^{iz\sin(x)} dx$, we obtain 
\begin{equation} \label{eq:G}
\begin{split}
&G(t_1, t_2) = (2\pi)^D e^{-i(\mathcal{E}_g + \Delta)(t_2-t_1)} B(|t_2 - t_1|) \prod_{k=1}^D J_0\left(|\Phi_k|\right).
\end{split}
\end{equation}

Unlike the Keldysh formalism, which integrates over the time in Eq. \eqref{eq:L} assuming a continuous-wave field, our approach does not use any assumption on the shape or the pulse width of the laser field, yielding Eqs. \eqref{eq:Wfinal}--\eqref{eq:G}, which allow the CB population to be calculated for a laser field of an arbitrary waveform and pulse width. The Keldysh theory calculates the $L(p)$ amplitude, in accordance with Eq. \eqref{eq:L}, at the first step, followed by integration over the momentum, as prescribed by Eq. \eqref{eq:W}, thus yielding the field-cycle-averaged ionization rate for a dielectric with a Kane-type dispersion in the presence of a cw laser field. Our approach, on the other hand, integrates over the momentum at the first step for a periodic dispersion relation, which is better suited for the strong-field regime. This procedure yields the two-time $G(t_1, t_2)$ ionization cross-section function, which is used in the second step to calculate, through the integration over the time, the CB population for a laser field of arbitrary waveform and pulse width.

\begin{figure} 
\includegraphics[width=0.5\textwidth]{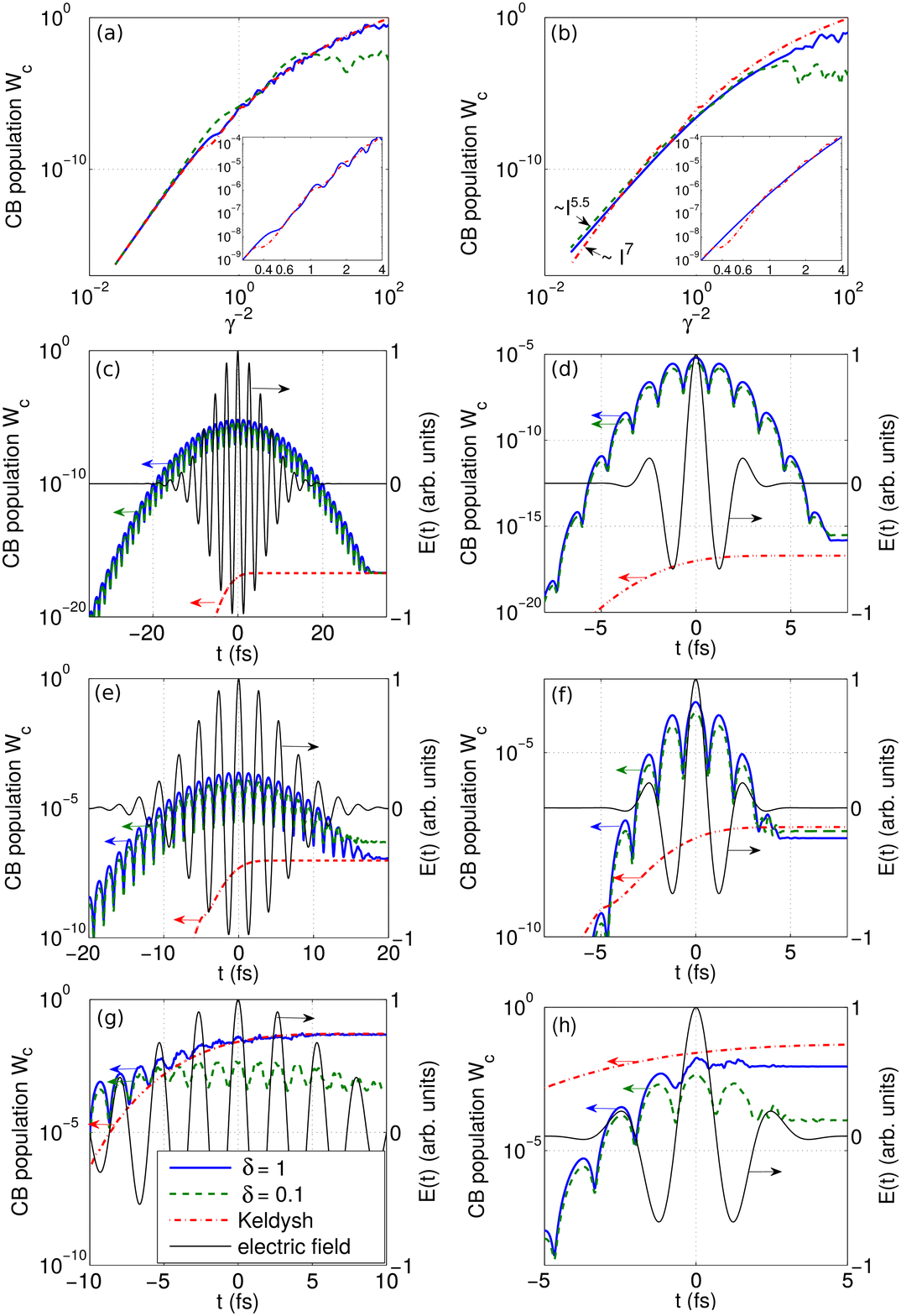} 
\caption{\label{fig:a1}(color online) (a),(b) The CB population in the wake of the pulse calculated as a function of $\gamma^{-2}$ with the use of Eqs. \eqref{eq:Wfinal}--\eqref{eq:G} with $\delta = 1$ (solid blue line), with $\delta = 0.1$ (dashed green line) and using the Keldysh formula (dash-dotted red line). The insets show a close-up of Franz--Keldysh modulation.  (c)--(h) Dynamics of the CB population calculated with the use of Eqs.  \eqref{eq:Wfinal}, \eqref{eq:G} with $\delta = 1$(solid blue line) and $\delta = 0.1$ (dashed green line) and using the Keldysh formula (dash-dotted red line) for $\gamma$ = 7.2 (c),(d), $\gamma$ = 1.2 (e),(f) and $\gamma$ = 0.2 (g),(h). The driver field is shown by the thin black line (right axis). The FWHM pulse width is 10 fs (a),(c),(e),(g), 2.4 fs (b),(d),(f),(h), the central wavelength of the driver field is $\lambda_0 = 800$ nm, $\mathcal{E}_g/\omega = 6.45$, $B(\tau)\equiv1$, and $D=1$.  The color arrows relate the curves of the same color to the respective (left or right) ordinate axis.
 }
\end{figure}

Unlike the periodic dispersion relation of Eq. \eqref{eq:cosine_bandD}, the Kane-type band model, used in the Keldysh treatment, is not suited to describe the dispersion near the zone edges. Predictions of Eqs. \eqref{eq:Wfinal}--\eqref{eq:G} can therefore agree with the Keldysh formula only for relatively low field intensities, where $|x| < \pi/2$, so that the dispersion relation of Eq. $\eqref{eq:cosine_bandD}$ can be approximated by a second-order Taylor-series polynomial. In terms of the Keldysh adiabaticity parameter, $\gamma = \frac{\omega \sqrt{m \Delta}}{E_0}$, where $\omega$ is the field frequency, and $E_0$ is the field amplitude, this condition is written as $\gamma > \frac{2}{\pi\sqrt{\delta}}$, where $\delta = \Delta/E_g$. Furthermore, since the Keldysh formula was derived for a cw field, discrepancies between the predictions of Eqs. \eqref{eq:Wfinal} -- \eqref{eq:G} and the Keldysh formula are expected to grow for shorter pulse widths.

Results of calculations for a one-dimensional decoherence-free semiconductor fully justify these expectations [Figs. \ref{fig:a1}(a) -- \ref{fig:a1}(h)]. For relatively long pulse widths and low field intensities [Figs. \ref{fig:a1}(a),(c),(e)], when both $\gamma > 1$ and $\gamma > \frac{2}{\pi\sqrt{\delta}}$ conditions are satisfied, Eqs. \eqref{eq:Wfinal} -- \eqref{eq:G} are seen to accurately reproduce the $I^k$ scaling as an asymptotic behavior for the CB population in the wake of the laser pulse as a function of the field intensity $I$, with $k$ being the minimum number of photons needed to surpass the band gap. However, CB population dynamics within each field cycle, as calculations using Eqs. \eqref{eq:Wfinal} - \eqref{eq:G} show, can drastically differ from predictions of the Keldysh formula even in the case of sufficiently long pulse widths and $\gamma > \frac{2}{\pi\sqrt{\delta}}$ [Figs. \ref{fig:a1}(c),\ref{fig:a1}(e)]. Specifically, in the regime of low field intensities [Figs. \ref{fig:a1}(c)--\ref{fig:a1}(f)], the CB population displays a pronounced oscillatory behavior, following the cycles of the laser field \cite{Schiffrin2013}. This oscillatory dynamics within the field half-cycle shows that, in the regime of low field intensities, most of the population transferred from the valence to the conduction band returns back to the valence band within the same field half-cycle. When the driver pulse is long enough, however, this oscillatory dynamics converges to the Keldysh theory result in the wake of the laser pulse [Figs. \ref{fig:a1}(b), \ref{fig:a1}(c)], indicating the buildup of the multiphoton regime of photoionization as an asymptotic behavior of CB population. Moreover, Eqs. \eqref{eq:Wfinal} - \eqref{eq:G} are seen to accurately reproduce stepwise changes in the CB population as a function of the field intensity [the $\gamma^{-2}$ parameter in Fig. \ref{fig:a1}(a)] due to the Franz--Keldysh modulation \cite{Keldysh1958, Franz1958} of the band gap [see the inset in Fig. \ref{fig:a1}(a)].

It is clearly seen from Fig. \ref{fig:a1}(a) that the CB population in the wake of the laser pulse calculated with the use of Eqs. \eqref{eq:Wfinal}--\eqref{eq:G} as a function of $\gamma^{-2}$ (i.e., parameter proportional to the field intensity $I$) closely follows predictions of the Keldysh theory for $\gamma > \frac{2}{\pi\sqrt{\delta}}$, but noticeably deviates from the Keldysh theory result when this inequality is not satisfied [e.g., for $\gamma > 2$ in the case of $\delta = 0.1$ in Fig. \ref{fig:a1}(a)].

In the case of very short laser pulses, where each field half-cycle significantly differs in its intensity from the adjacent field half-cycles [thin black line in Figs. \ref{fig:a1}(d), \ref{fig:a1}(f), \ref{fig:a1}(h)], the integration over time in Eqs. \eqref{eq:Wfinal} and \eqref{eq:W} no longer converges to the Keldysh theory result even in the wake of the pulse [Figs. \ref{fig:a1}(b), \ref{fig:a1}(d), \ref{fig:a1}(f), \ref{fig:a1}(h)]. Because the number of photons needed for ionization is no longer defined in the regime of very short light pulses, the Franz--Keldysh modulation of the CB population as a function of the field intensity is much less pronounced and is not observed where predicted by the Keldysh formula for a cw field [the inset in Fig. \ref{fig:a1}(b)].

In the high-intensity regime, $\gamma < 1$, the CB population rapidly builds up after each field half-cycle, giving rise to a stepwise growth of the CB electron density [Figs. \ref{fig:a1}(g), \ref{fig:a1}(h)]. 
Because of a rapidly oscillating factor under the integral, $\Phi$ is vanishingly small unless $|t_2 - t_1| < \epsilon^{-1}$, where $\epsilon = d_1 E$, and $d_1$ is the lattice constant. We can therefore use a power-series expansion $\mu(\tau) = \mu(t_2) -\epsilon(t_2-\tau)$ to reduce the expression for $G(t_1, t_2)$ to find in a 1D case
\begin{equation} 
G(t_1, t_2) = 2\pi {E^2} e^{-2i\frac{\mathcal{E}_g + \Delta}{|\epsilon|}\xi} J_0\left(\frac{2\Delta}{|\epsilon|}|\sin \xi| \right),
\end{equation}
where $\xi = \frac{|\epsilon|}{2}(t_2 - t_1)$. The ionization rate can be then written as
\begin{equation}
w(E)=\frac{dW_{c}}{dt} = 4 \frac{
|E|}{d_1} |\mathcal{{N}}|^2 \mathrm{Re} \int_{-\pi}^{\pi} d\eta \int_{-\infty}^{\infty} d\xi e^{-is\nu(\xi,\eta)}, \label{eq:wE}
\end{equation}
where $s = 2\frac{\Delta}{|\epsilon|}$ and $\nu(\xi,\eta) = (\delta^{-1} + 1)\xi + \sin \xi \sin \eta$.

To simultaneously satisfy the inequalities $\frac{2}{\pi\sqrt{\delta}} < \gamma < 1$, we require $\delta > (\frac{\pi}{2})^2$ and calculate the integrals in Eq. \eqref{eq:wE} using the saddle-point method to derive
in the first order in  $\delta^{-1/2}$, we obtain ($K$ is a constant):
\begin{equation} \label{eq:wtun}
w(E) = {K}E^2e^{-\frac{4}{3}\sqrt{\frac{2}{\delta}}\frac{\mathcal{E}_g}{|\epsilon|}} = {K}E^2e^{-\frac{4}{3}\frac{(2m)^{1/2} \mathcal{E}_g^{3/2}}{|E|}},
\end{equation}
where K is the field-independent numerical factor.

Eq. \eqref{eq:wtun} recovers not only the signature tunneling exponential, but also the $E^2$ scaling of the pre-exponential factor \cite{Keldysh1958, Kane1959}.

\begin{figure} 
\includegraphics[width=0.5\textwidth]{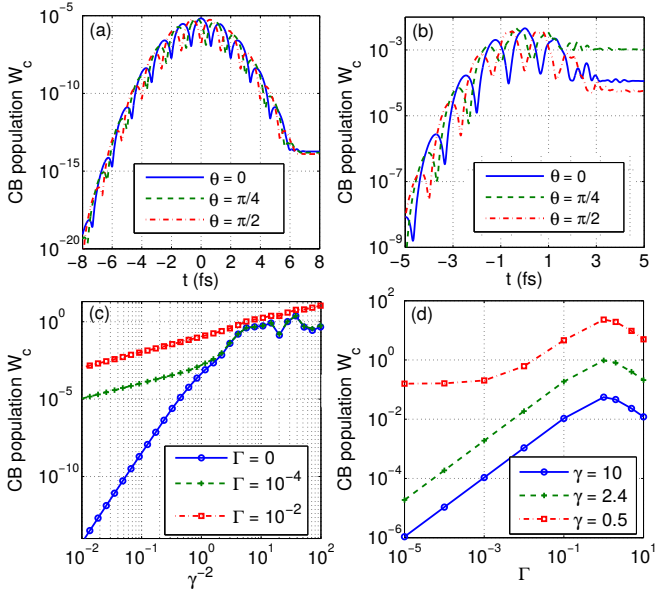}
\caption{\label{fig:W_cep}(color online) Dynamics of the CB population for $\gamma = 5$ (a) and 0.2 (b) calculated using Eqs. \eqref{eq:Wfinal},\eqref{eq:G} for the CEP $\theta = 0$ (dashed blue line), $\pi/4$ (solid green line), $\pi/2$ (dash-dotted red line). (c),(d) CB population in the wake of the pulse (b) as a function of $\gamma^{-2}$ for different $\Gamma$ and (d) as a function of $\Gamma$ for different $\gamma$.  The FWHM pulse width is 2.4 fs, $\lambda_0 = 800$ nm, $\mathcal{E}_g/\omega = 6.45$, $\delta = 0.1$, $D=1$.}
\end{figure}

To understand effects related to the carrier-envelope phase (CEP), we represent the driver field as $E(t) = E_0 e^{-(t/T)^2} \cos\left(\omega t + \theta\right)$, where $T$ is the pulse duration, and examine the CB population as a function of the CEP $\theta$. In the case of long laser pulses, containing many field cycles, i.e., in the regime where Eqs. \eqref{eq:Wfinal}--\eqref{eq:G} recover the results of the Keldysh theory for cw fields, no CEP dependence is observed, in full agreement with the Keldysh theory.
For very short laser pulses of low intensity, the instantaneous CB population within the field half-cycle is sensitive to the CEP [Fig. \ref{fig:W_cep}(a)]. However, the CB population left in the wake of the driver pulse is virtually CEP-independent [$t > 6$ fs in Fig. \ref{fig:W_cep}(a)], with almost no deviation from the Keldysh theory. In the regime of high field intensities [Fig. \ref{fig:W_cep}(b)], the CB density in the wake of the pulse can be represented as a sum of populations transferred to the conduction band by each field half-cycle [Fig. \ref{fig:W_cep}(b)]. The CB population induced by a single field half-cycle, in its turn, is a strongly nonlinear function of the field intensity achieved within this half-cycle. As a result, the CB population in the wake of a very short driver pulse is efficiently controlled by the CEP of this pulse, changing by an order of magnitude in Fig. \ref{fig:W_cep}(b) as the CEP is shifted by $\pi/4$.

Decoherence effects, which can be included in the model through the $B(\tau)$ factor in Eq. \eqref{eq:G}, lead to a gradual loss of phase memory in photoinization. Using a phenomenological ansatz $B(\tau) = e^{-\Gamma \mathcal{E}_g \tau}$, with decoherence constant $\Gamma$, defining the coherence time as $\tau_c = (\Gamma \mathcal{E}_g )^{-1}$, we find that changes in photoionization are especially dramatic in the low-intensity regime [Figs. \ref{fig:W_cep}(c), \ref{fig:W_cep}(d)], where the CB population left in the wake of the pulse is controlled by the interference of electron wave packets induced by each field half-cycle [Figs. \ref{fig:a1}(c),\ref{fig:a1}(d), \ref{fig:W_cep}(a), \ref{fig:W_cep}(b)]. In this regime, decoherence effects tend to prevent a coherent cancellation of the ionization probability within each field half-cycle [Figs. \ref{fig:a1}(c),\ref{fig:a1}(d), \ref{fig:W_cep}(a), \ref{fig:W_cep}(b)], increasing the CB population in the wake of the pulse (Figs. \ref{fig:W_cep}(c), \ref{fig:W_cep}(d)) and giving rise to deviations from the $I^n$ scaling of the ionization rate, which would be typical of $n$-photon ionization in the absence of decoherence. As decoherence becomes stronger, the intensity dependence of the ionization rate coverges to the $I$ scaling [Fig. \ref{fig:W_cep}(c)]. Strong decoherence can also suppress the coherent buildup of the CB population within each field half-cycle. This effect is clearly seen in Fig. \ref{fig:W_cep}(d), where the CB population in the wake of the pulse starts to decrease with increasing $\Gamma$ as $\tau_c$ becomes shorter than $\mathcal{E}_g^{-1}$.

To summarize, we have extended the Keldysh theory of photoionzation of semiconductors to the case of ultrashort driver pulses of arbitrary waveform and pulse width. We derived a closed-form solution for the nonadiabatic ionization rate in a transparent solid, which can be used not only to calculate the probability of ionization in the wake of the pulse, but also to examine ultrafast ionization dynamics within the field cycle. Our approach has been shown to accurately recover the results of the Keldysh theory within its range of applicability.   

This research was supported in part by the Russian Foundation for Basic Research (project nos. 13-02-01465, 13-02-92115,  14-02-90030), the Welch Foundation (Grant No. A-1801), and the Russian Science Foundation (project no. 14-12-00772).
\bibliography{lib}

\begin{thebibliography}{27}%
\makeatletter
\providecommand \@ifxundefined [1]{%
 \@ifx{#1\undefined}
}%
\providecommand \@ifnum [1]{%
 \ifnum #1\expandafter \@firstoftwo
 \else \expandafter \@secondoftwo
 \fi
}%
\providecommand \@ifx [1]{%
 \ifx #1\expandafter \@firstoftwo
 \else \expandafter \@secondoftwo
 \fi
}%
\providecommand \natexlab [1]{#1}%
\providecommand \enquote  [1]{``#1''}%
\providecommand \bibnamefont  [1]{#1}%
\providecommand \bibfnamefont [1]{#1}%
\providecommand \citenamefont [1]{#1}%
\providecommand \href@noop [0]{\@secondoftwo}%
\providecommand \href [0]{\begingroup \@sanitize@url \@href}%
\providecommand \@href[1]{\@@startlink{#1}\@@href}%
\providecommand \@@href[1]{\endgroup#1\@@endlink}%
\providecommand \@sanitize@url [0]{\catcode `\\12\catcode `\$12\catcode
  `\&12\catcode `\#12\catcode `\^12\catcode `\_12\catcode `\%12\relax}%
\providecommand \@@startlink[1]{}%
\providecommand \@@endlink[0]{}%
\providecommand \url  [0]{\begingroup\@sanitize@url \@url }%
\providecommand \@url [1]{\endgroup\@href {#1}{\urlprefix }}%
\providecommand \urlprefix  [0]{URL }%
\providecommand \Eprint [0]{\href }%
\providecommand \doibase [0]{http://dx.doi.org/}%
\providecommand \selectlanguage [0]{\@gobble}%
\providecommand \bibinfo  [0]{\@secondoftwo}%
\providecommand \bibfield  [0]{\@secondoftwo}%
\providecommand \translation [1]{[#1]}%
\providecommand \BibitemOpen [0]{}%
\providecommand \bibitemStop [0]{}%
\providecommand \bibitemNoStop [0]{.\EOS\space}%
\providecommand \EOS [0]{\spacefactor3000\relax}%
\providecommand \BibitemShut  [1]{\csname bibitem#1\endcsname}%
\let\auto@bib@innerbib\@empty
\bibitem [{\citenamefont {Keldysh}(1965)}]{Keldysh1965}%
  \BibitemOpen
  \bibfield  {author} {\bibinfo {author} {\bibfnamefont {L.~V.}\ \bibnamefont
  {Keldysh}},\ }\href@noop {} {\bibfield  {journal} {\bibinfo  {journal} {Sov.
  Phys. JETP}\ }\textbf {\bibinfo {volume} {20}},\ \bibinfo {pages} {1307}
  (\bibinfo {year} {1965})}\BibitemShut {NoStop}%
\bibitem [{\citenamefont {Bloembergen}(1974)}]{Bloembergen1974}%
  \BibitemOpen
  \bibfield  {author} {\bibinfo {author} {\bibfnamefont {N.}~\bibnamefont
  {Bloembergen}},\ }\href@noop {} {\bibfield  {journal} {\bibinfo  {journal}
  {IEEE J. Quantum Electron.}\ }\textbf {\bibinfo {volume} {10}},\ \bibinfo
  {pages} {375} (\bibinfo {year} {1974})}\BibitemShut {NoStop}%
\bibitem [{\citenamefont {Lenzner}\ \emph {et~al.}(1998)\citenamefont
  {Lenzner}, \citenamefont {Kr\"{u}ger}, \citenamefont {Sartania},
  \citenamefont {Cheng}, \citenamefont {Spielmann}, \citenamefont {Mourou},
  \citenamefont {Kautek},\ and\ \citenamefont {Krausz}}]{Lenzner1998}%
  \BibitemOpen
  \bibfield  {author} {\bibinfo {author} {\bibfnamefont {M.}~\bibnamefont
  {Lenzner}}, \bibinfo {author} {\bibfnamefont {J.}~\bibnamefont {Kr\"{u}ger}},
  \bibinfo {author} {\bibfnamefont {S.}~\bibnamefont {Sartania}}, \bibinfo
  {author} {\bibfnamefont {Z.}~\bibnamefont {Cheng}}, \bibinfo {author}
  {\bibfnamefont {C.}~\bibnamefont {Spielmann}}, \bibinfo {author}
  {\bibfnamefont {G.}~\bibnamefont {Mourou}}, \bibinfo {author} {\bibfnamefont
  {W.}~\bibnamefont {Kautek}}, \ and\ \bibinfo {author} {\bibfnamefont
  {F.}~\bibnamefont {Krausz}},\ }\href@noop {} {\bibfield  {journal} {\bibinfo
  {journal} {Phys. Rev. Lett.}\ }\textbf {\bibinfo {volume} {80}},\ \bibinfo
  {pages} {4076} (\bibinfo {year} {1998})}\BibitemShut {NoStop}%
\bibitem [{\citenamefont {Brabec}\ and\ \citenamefont
  {Krausz}(2000)}]{Brabec2000}%
  \BibitemOpen
  \bibfield  {author} {\bibinfo {author} {\bibfnamefont {T.}~\bibnamefont
  {Brabec}}\ and\ \bibinfo {author} {\bibfnamefont {F.}~\bibnamefont
  {Krausz}},\ }\href {\doibase 10.1103/RevModPhys.72.545} {\bibfield  {journal}
  {\bibinfo  {journal} {Rev. Mod. Phys.}\ }\textbf {\bibinfo {volume} {72}},\
  \bibinfo {pages} {545} (\bibinfo {year} {2000})}\BibitemShut {NoStop}%
\bibitem [{\citenamefont {Tonouchi}(2002)}]{Tonouchi2002}%
  \BibitemOpen
  \bibfield  {author} {\bibinfo {author} {\bibfnamefont {M.}~\bibnamefont
  {Tonouchi}},\ }\href@noop {} {\bibfield  {journal} {\bibinfo  {journal} {Nat.
  Photonics}\ }\textbf {\bibinfo {volume} {1}},\ \bibinfo {pages} {97}
  (\bibinfo {year} {2002})}\BibitemShut {NoStop}%
\bibitem [{\citenamefont {Couairon}\ and\ \citenamefont
  {Mysyrowicz}(2007)}]{Couairon2007}%
  \BibitemOpen
  \bibfield  {author} {\bibinfo {author} {\bibfnamefont {A.}~\bibnamefont
  {Couairon}}\ and\ \bibinfo {author} {\bibfnamefont {A.}~\bibnamefont
  {Mysyrowicz}},\ }\href {\doibase 10.1016/j.physrep.2006.12.005} {\bibfield
  {journal} {\bibinfo  {journal} {Phys. Rep.}\ }\textbf {\bibinfo {volume}
  {441}},\ \bibinfo {pages} {47} (\bibinfo {year} {2007})}\BibitemShut
  {NoStop}%
\bibitem [{\citenamefont {Berg\'{e}}\ \emph {et~al.}(2007)\citenamefont
  {Berg\'{e}}, \citenamefont {Skupin}, \citenamefont {Nuter}, \citenamefont
  {Kasparian},\ and\ \citenamefont {Wolf}}]{Berge2007}%
  \BibitemOpen
  \bibfield  {author} {\bibinfo {author} {\bibfnamefont {L.}~\bibnamefont
  {Berg\'{e}}}, \bibinfo {author} {\bibfnamefont {S.}~\bibnamefont {Skupin}},
  \bibinfo {author} {\bibfnamefont {R.}~\bibnamefont {Nuter}}, \bibinfo
  {author} {\bibfnamefont {J.}~\bibnamefont {Kasparian}}, \ and\ \bibinfo
  {author} {\bibfnamefont {J.-P.}\ \bibnamefont {Wolf}},\ }\href {\doibase
  10.1088/0034-4885/70/10/R03} {\bibfield  {journal} {\bibinfo  {journal}
  {Reports Prog. Phys.}\ }\textbf {\bibinfo {volume} {70}},\ \bibinfo {pages}
  {1633} (\bibinfo {year} {2007})}\BibitemShut {NoStop}%
\bibitem [{\citenamefont {Perelomov}\ \emph {et~al.}(1966)\citenamefont
  {Perelomov}, \citenamefont {Popov},\ and\ \citenamefont
  {Terent'ev}}]{Perelomov1966}%
  \BibitemOpen
  \bibfield  {author} {\bibinfo {author} {\bibfnamefont {A.~M.}\ \bibnamefont
  {Perelomov}}, \bibinfo {author} {\bibfnamefont {V.~S.}\ \bibnamefont
  {Popov}}, \ and\ \bibinfo {author} {\bibfnamefont {M.~V.}\ \bibnamefont
  {Terent'ev}},\ }\href@noop {} {\bibfield  {journal} {\bibinfo  {journal}
  {Sov. Phys. JETP}\ }\textbf {\bibinfo {volume} {23}},\ \bibinfo {pages} {924}
  (\bibinfo {year} {1966})}\BibitemShut {NoStop}%
\bibitem [{\citenamefont {Ammosov}\ \emph {et~al.}(1986)\citenamefont
  {Ammosov}, \citenamefont {Delone},\ and\ \citenamefont
  {Krainov}}]{Ammosov1986}%
  \BibitemOpen
  \bibfield  {author} {\bibinfo {author} {\bibfnamefont {M.~V.}\ \bibnamefont
  {Ammosov}}, \bibinfo {author} {\bibfnamefont {N.~B.}\ \bibnamefont {Delone}},
  \ and\ \bibinfo {author} {\bibfnamefont {V.~P.}\ \bibnamefont {Krainov}},\
  }\href
  {http://www.mendeley.com/research/tunnel-ionization-of-complex-atoms-and-of-atomic-ions-in-an-alternating-electromagnetic-field-2/}
  {\bibfield  {journal} {\bibinfo  {journal} {Sov Phys JETP}\ }\textbf
  {\bibinfo {volume} {64}},\ \bibinfo {pages} {4} (\bibinfo {year}
  {1986})}\BibitemShut {NoStop}%
\bibitem [{\citenamefont {Yudin}\ and\ \citenamefont
  {Ivanov}(2001)}]{Yudin2001}%
  \BibitemOpen
  \bibfield  {author} {\bibinfo {author} {\bibfnamefont {G.}~\bibnamefont
  {Yudin}}\ and\ \bibinfo {author} {\bibfnamefont {M.}~\bibnamefont {Ivanov}},\
  }\href {\doibase 10.1103/PhysRevA.64.013409} {\bibfield  {journal} {\bibinfo
  {journal} {Phys. Rev. A}\ }\textbf {\bibinfo {volume} {64}},\ \bibinfo
  {pages} {6} (\bibinfo {year} {2001})}\BibitemShut {NoStop}%
\bibitem [{\citenamefont {Goulielmakis}\ \emph {et~al.}(2007)\citenamefont
  {Goulielmakis}, \citenamefont {Yakovlev}, \citenamefont {Cavalieri},
  \citenamefont {Uiberacker}, \citenamefont {Pervak}, \citenamefont
  {Apolonski}, \citenamefont {Kienberger}, \citenamefont {Kleineberg},\ and\
  \citenamefont {Krausz}}]{Goulielmakis2007}%
  \BibitemOpen
  \bibfield  {author} {\bibinfo {author} {\bibfnamefont {E.}~\bibnamefont
  {Goulielmakis}}, \bibinfo {author} {\bibfnamefont {V.~S.}\ \bibnamefont
  {Yakovlev}}, \bibinfo {author} {\bibfnamefont {A.~L.}\ \bibnamefont
  {Cavalieri}}, \bibinfo {author} {\bibfnamefont {M.}~\bibnamefont
  {Uiberacker}}, \bibinfo {author} {\bibfnamefont {V.}~\bibnamefont {Pervak}},
  \bibinfo {author} {\bibfnamefont {A.}~\bibnamefont {Apolonski}}, \bibinfo
  {author} {\bibfnamefont {R.}~\bibnamefont {Kienberger}}, \bibinfo {author}
  {\bibfnamefont {U.}~\bibnamefont {Kleineberg}}, \ and\ \bibinfo {author}
  {\bibfnamefont {F.}~\bibnamefont {Krausz}},\ }\href {\doibase
  10.1126/science.1142855} {\bibfield  {journal} {\bibinfo  {journal}
  {Science}\ }\textbf {\bibinfo {volume} {317}},\ \bibinfo {pages} {769}
  (\bibinfo {year} {2007})}\BibitemShut {NoStop}%
\bibitem [{\citenamefont {Corkum}\ and\ \citenamefont
  {Krausz}(2007)}]{Corkum2007}%
  \BibitemOpen
  \bibfield  {author} {\bibinfo {author} {\bibfnamefont {P.~B.}\ \bibnamefont
  {Corkum}}\ and\ \bibinfo {author} {\bibfnamefont {F.}~\bibnamefont
  {Krausz}},\ }\href
  {http://www.nature.com/nphys/journal/v3/n6/pdf/nphys620.pdf} {\bibfield
  {journal} {\bibinfo  {journal} {Nat. Phys.}\ }\textbf {\bibinfo {volume}
  {3}},\ \bibinfo {pages} {381} (\bibinfo {year} {2007})}\BibitemShut {NoStop}%
\bibitem [{\citenamefont {Uiberacker}\ \emph {et~al.}(2007)\citenamefont
  {Uiberacker}, \citenamefont {Uphues}, \citenamefont {Schultze}, \citenamefont
  {Verhoef}, \citenamefont {Yakovlev}, \citenamefont {Kling}, \citenamefont
  {Rauschenberger}, \citenamefont {Kabachnik}, \citenamefont {Schr\"{o}der},
  \citenamefont {Lezius}, \citenamefont {Kompa}, \citenamefont {Muller},
  \citenamefont {Vrakking}, \citenamefont {Hendel}, \citenamefont {Kleineberg},
  \citenamefont {Heinzmann}, \citenamefont {Drescher},\ and\ \citenamefont
  {Krausz}}]{Uiberacker2007}%
  \BibitemOpen
  \bibfield  {author} {\bibinfo {author} {\bibfnamefont {M.}~\bibnamefont
  {Uiberacker}}, \bibinfo {author} {\bibfnamefont {T.}~\bibnamefont {Uphues}},
  \bibinfo {author} {\bibfnamefont {M.}~\bibnamefont {Schultze}}, \bibinfo
  {author} {\bibfnamefont {A.~J.}\ \bibnamefont {Verhoef}}, \bibinfo {author}
  {\bibfnamefont {V.}~\bibnamefont {Yakovlev}}, \bibinfo {author}
  {\bibfnamefont {M.~F.}\ \bibnamefont {Kling}}, \bibinfo {author}
  {\bibfnamefont {J.}~\bibnamefont {Rauschenberger}}, \bibinfo {author}
  {\bibfnamefont {N.~M.}\ \bibnamefont {Kabachnik}}, \bibinfo {author}
  {\bibfnamefont {H.}~\bibnamefont {Schr\"{o}der}}, \bibinfo {author}
  {\bibfnamefont {M.}~\bibnamefont {Lezius}}, \bibinfo {author} {\bibfnamefont
  {K.~L.}\ \bibnamefont {Kompa}}, \bibinfo {author} {\bibfnamefont {H.-G.}\
  \bibnamefont {Muller}}, \bibinfo {author} {\bibfnamefont {M.~J.~J.}\
  \bibnamefont {Vrakking}}, \bibinfo {author} {\bibfnamefont {S.}~\bibnamefont
  {Hendel}}, \bibinfo {author} {\bibfnamefont {U.}~\bibnamefont {Kleineberg}},
  \bibinfo {author} {\bibfnamefont {U.}~\bibnamefont {Heinzmann}}, \bibinfo
  {author} {\bibfnamefont {M.}~\bibnamefont {Drescher}}, \ and\ \bibinfo
  {author} {\bibfnamefont {F.}~\bibnamefont {Krausz}},\ }\href {\doibase
  10.1038/nature05648} {\bibfield  {journal} {\bibinfo  {journal} {Nature}\
  }\textbf {\bibinfo {volume} {446}},\ \bibinfo {pages} {627} (\bibinfo {year}
  {2007})}\BibitemShut {NoStop}%
\bibitem [{\citenamefont {Balciunas}\ \emph {et~al.}(2013)\citenamefont
  {Balciunas}, \citenamefont {Verhoef}, \citenamefont {Mitrofanov},
  \citenamefont {Fan}, \citenamefont {Serebryannikov}, \citenamefont {Ivanov},
  \citenamefont {Zheltikov},\ and\ \citenamefont {Baltuska}}]{Balciunas2013}%
  \BibitemOpen
  \bibfield  {author} {\bibinfo {author} {\bibfnamefont {T.}~\bibnamefont
  {Balciunas}}, \bibinfo {author} {\bibfnamefont {A.~J.}\ \bibnamefont
  {Verhoef}}, \bibinfo {author} {\bibfnamefont {A.~V.}\ \bibnamefont
  {Mitrofanov}}, \bibinfo {author} {\bibfnamefont {G.}~\bibnamefont {Fan}},
  \bibinfo {author} {\bibfnamefont {E.~E.}\ \bibnamefont {Serebryannikov}},
  \bibinfo {author} {\bibfnamefont {M.~Y.}\ \bibnamefont {Ivanov}}, \bibinfo
  {author} {\bibfnamefont {A.~M.}\ \bibnamefont {Zheltikov}}, \ and\ \bibinfo
  {author} {\bibfnamefont {A.}~\bibnamefont {Baltuska}},\ }\href {\doibase
  10.1016/j.chemphys.2012.02.007} {\bibfield  {journal} {\bibinfo  {journal}
  {Chem. Phys.}\ }\textbf {\bibinfo {volume} {414}},\ \bibinfo {pages} {92}
  (\bibinfo {year} {2013})}\BibitemShut {NoStop}%
\bibitem [{\citenamefont {Verhoef}\ \emph {et~al.}(2010)\citenamefont
  {Verhoef}, \citenamefont {Mitrofanov}, \citenamefont {Serebryannikov},
  \citenamefont {Kartashov}, \citenamefont {Zheltikov},\ and\ \citenamefont
  {Baltu\v{s}ka}}]{Verhoef2010}%
  \BibitemOpen
  \bibfield  {author} {\bibinfo {author} {\bibfnamefont {A.~J.}\ \bibnamefont
  {Verhoef}}, \bibinfo {author} {\bibfnamefont {A.~V.}\ \bibnamefont
  {Mitrofanov}}, \bibinfo {author} {\bibfnamefont {E.~E.}\ \bibnamefont
  {Serebryannikov}}, \bibinfo {author} {\bibfnamefont {D.~V.}\ \bibnamefont
  {Kartashov}}, \bibinfo {author} {\bibfnamefont {A.~M.}\ \bibnamefont
  {Zheltikov}}, \ and\ \bibinfo {author} {\bibfnamefont {A.}~\bibnamefont
  {Baltu\v{s}ka}},\ }\href {\doibase 10.1103/PhysRevLett.104.163904} {\bibfield
   {journal} {\bibinfo  {journal} {Phys. Rev. Lett.}\ }\textbf {\bibinfo
  {volume} {104}},\ \bibinfo {pages} {1} (\bibinfo {year} {2010})}\BibitemShut
  {NoStop}%
\bibitem [{\citenamefont {Mitrofanov}\ \emph {et~al.}(2011)\citenamefont
  {Mitrofanov}, \citenamefont {Verhoef}, \citenamefont {Serebryannikov},
  \citenamefont {Lumeau}, \citenamefont {Glebov}, \citenamefont {Zheltikov},\
  and\ \citenamefont {Baltu\v{s}ka}}]{Mitrofanov2011}%
  \BibitemOpen
  \bibfield  {author} {\bibinfo {author} {\bibfnamefont {A.~V.}\ \bibnamefont
  {Mitrofanov}}, \bibinfo {author} {\bibfnamefont {A.~J.}\ \bibnamefont
  {Verhoef}}, \bibinfo {author} {\bibfnamefont {E.~E.}\ \bibnamefont
  {Serebryannikov}}, \bibinfo {author} {\bibfnamefont {J.}~\bibnamefont
  {Lumeau}}, \bibinfo {author} {\bibfnamefont {L.}~\bibnamefont {Glebov}},
  \bibinfo {author} {\bibfnamefont {A.~M.}\ \bibnamefont {Zheltikov}}, \ and\
  \bibinfo {author} {\bibfnamefont {A.}~\bibnamefont {Baltu\v{s}ka}},\ }\href
  {\doibase 10.1103/PhysRevLett.106.147401} {\bibfield  {journal} {\bibinfo
  {journal} {Phys. Rev. Lett.}\ }\textbf {\bibinfo {volume} {106}},\ \bibinfo
  {pages} {147401} (\bibinfo {year} {2011})}\BibitemShut {NoStop}%
\bibitem [{\citenamefont {Kane}(1959)}]{Kane1959}%
  \BibitemOpen
  \bibfield  {author} {\bibinfo {author} {\bibfnamefont {E.~O.}\ \bibnamefont
  {Kane}},\ }\href@noop {} {\bibfield  {journal} {\bibinfo  {journal} {J. Phys.
  Chem. Solids}\ }\textbf {\bibinfo {volume} {12}},\ \bibinfo {pages} {181}
  (\bibinfo {year} {1959})}\BibitemShut {NoStop}%
\bibitem [{\citenamefont {Kane}(1961)}]{Kane1961}%
  \BibitemOpen
  \bibfield  {author} {\bibinfo {author} {\bibfnamefont {E.~O.}\ \bibnamefont
  {Kane}},\ }\href {\doibase 10.1063/1.1735965} {\bibfield  {journal} {\bibinfo
   {journal} {J. Appl. Phys.}\ }\textbf {\bibinfo {volume} {32}},\ \bibinfo
  {pages} {83} (\bibinfo {year} {1961})}\BibitemShut {NoStop}%
\bibitem [{\citenamefont {{Bonch-Bruevich, V. L.
  Kalaschnikov}}(1982)}]{Bonch-Bruevich1982}%
  \BibitemOpen
  \bibfield  {author} {\bibinfo {author} {\bibfnamefont {S.~G.}\ \bibnamefont
  {{Bonch-Bruevich, V. L. Kalaschnikov}}},\ }\href@noop {} {\emph {\bibinfo
  {title} {{Halbleiterphysik}}}}\ (\bibinfo  {publisher} {VEB},\ \bibinfo
  {address} {Berlin},\ \bibinfo {year} {1982})\BibitemShut {NoStop}%
\bibitem [{\citenamefont {Hawkins}\ and\ \citenamefont
  {Ivanov}(2013)}]{Hawkins2013}%
  \BibitemOpen
  \bibfield  {author} {\bibinfo {author} {\bibfnamefont {P.~G.}\ \bibnamefont
  {Hawkins}}\ and\ \bibinfo {author} {\bibfnamefont {M.~Y.}\ \bibnamefont
  {Ivanov}},\ }\href {\doibase 10.1103/PhysRevA.87.063842} {\bibfield
  {journal} {\bibinfo  {journal} {Phys. Rev. A}\ }\textbf {\bibinfo {volume}
  {87}},\ \bibinfo {pages} {063842} (\bibinfo {year} {2013})}\BibitemShut
  {NoStop}%
\bibitem [{\citenamefont {Schiffrin}\ \emph {et~al.}(2013)\citenamefont
  {Schiffrin}, \citenamefont {Paasch-Colberg}, \citenamefont {Karpowicz},
  \citenamefont {Apalkov}, \citenamefont {Gerster}, \citenamefont
  {M\"{u}hlbrandt}, \citenamefont {Korbman}, \citenamefont {Reichert},
  \citenamefont {Schultze}, \citenamefont {Holzner}, \citenamefont {Barth},
  \citenamefont {Kienberger}, \citenamefont {Ernstorfer}, \citenamefont
  {Yakovlev}, \citenamefont {Stockman},\ and\ \citenamefont
  {Krausz}}]{Schiffrin2013}%
  \BibitemOpen
  \bibfield  {author} {\bibinfo {author} {\bibfnamefont {A.}~\bibnamefont
  {Schiffrin}}, \bibinfo {author} {\bibfnamefont {T.}~\bibnamefont
  {Paasch-Colberg}}, \bibinfo {author} {\bibfnamefont {N.}~\bibnamefont
  {Karpowicz}}, \bibinfo {author} {\bibfnamefont {V.}~\bibnamefont {Apalkov}},
  \bibinfo {author} {\bibfnamefont {D.}~\bibnamefont {Gerster}}, \bibinfo
  {author} {\bibfnamefont {S.}~\bibnamefont {M\"{u}hlbrandt}}, \bibinfo
  {author} {\bibfnamefont {M.}~\bibnamefont {Korbman}}, \bibinfo {author}
  {\bibfnamefont {J.}~\bibnamefont {Reichert}}, \bibinfo {author}
  {\bibfnamefont {M.}~\bibnamefont {Schultze}}, \bibinfo {author}
  {\bibfnamefont {S.}~\bibnamefont {Holzner}}, \bibinfo {author} {\bibfnamefont
  {J.~V.}\ \bibnamefont {Barth}}, \bibinfo {author} {\bibfnamefont
  {R.}~\bibnamefont {Kienberger}}, \bibinfo {author} {\bibfnamefont
  {R.}~\bibnamefont {Ernstorfer}}, \bibinfo {author} {\bibfnamefont {V.~S.}\
  \bibnamefont {Yakovlev}}, \bibinfo {author} {\bibfnamefont {M.~I.}\
  \bibnamefont {Stockman}}, \ and\ \bibinfo {author} {\bibfnamefont
  {F.}~\bibnamefont {Krausz}},\ }\href {\doibase 10.1038/nature11567}
  {\bibfield  {journal} {\bibinfo  {journal} {Nature}\ }\textbf {\bibinfo
  {volume} {493}},\ \bibinfo {pages} {70} (\bibinfo {year} {2013})}\BibitemShut
  {NoStop}%
\bibitem [{\citenamefont {Serebryannikov}\ \emph {et~al.}(2009)\citenamefont
  {Serebryannikov}, \citenamefont {Verhoef}, \citenamefont {Mitrofanov},
  \citenamefont {Baltu\v{s}ka},\ and\ \citenamefont
  {Zheltikov}}]{Serebryannikov2009}%
  \BibitemOpen
  \bibfield  {author} {\bibinfo {author} {\bibfnamefont {E.}~\bibnamefont
  {Serebryannikov}}, \bibinfo {author} {\bibfnamefont {A.}~\bibnamefont
  {Verhoef}}, \bibinfo {author} {\bibfnamefont {A.}~\bibnamefont {Mitrofanov}},
  \bibinfo {author} {\bibfnamefont {A.}~\bibnamefont {Baltu\v{s}ka}}, \ and\
  \bibinfo {author} {\bibfnamefont {A.}~\bibnamefont {Zheltikov}},\ }\href
  {\doibase 10.1103/PhysRevA.80.053809} {\bibfield  {journal} {\bibinfo
  {journal} {Phys. Rev. A}\ }\textbf {\bibinfo {volume} {80}},\ \bibinfo
  {pages} {053809} (\bibinfo {year} {2009})}\BibitemShut {NoStop}%
\bibitem [{\citenamefont {Volkov}(1935)}]{Volkov1935}%
  \BibitemOpen
  \bibfield  {author} {\bibinfo {author} {\bibfnamefont {D.~M.}\ \bibnamefont
  {Volkov}},\ }\href@noop {} {\bibfield  {journal} {\bibinfo  {journal}
  {Zeitschrift fuer Phys.}\ }\textbf {\bibinfo {volume} {94}},\ \bibinfo
  {pages} {250} (\bibinfo {year} {1935})}\BibitemShut {NoStop}%
\bibitem [{\citenamefont {Kuehn}\ \emph {et~al.}(2010)\citenamefont {Kuehn},
  \citenamefont {Gaal}, \citenamefont {Reimann}, \citenamefont {Woerner},
  \citenamefont {Elsaesser},\ and\ \citenamefont {Hey}}]{Kuehn2010}%
  \BibitemOpen
  \bibfield  {author} {\bibinfo {author} {\bibfnamefont {W.}~\bibnamefont
  {Kuehn}}, \bibinfo {author} {\bibfnamefont {P.}~\bibnamefont {Gaal}},
  \bibinfo {author} {\bibfnamefont {K.}~\bibnamefont {Reimann}}, \bibinfo
  {author} {\bibfnamefont {M.}~\bibnamefont {Woerner}}, \bibinfo {author}
  {\bibfnamefont {T.}~\bibnamefont {Elsaesser}}, \ and\ \bibinfo {author}
  {\bibfnamefont {R.}~\bibnamefont {Hey}},\ }\href {\doibase
  10.1103/PhysRevB.82.075204} {\bibfield  {journal} {\bibinfo  {journal} {Phys.
  Rev. B}\ }\textbf {\bibinfo {volume} {82}},\ \bibinfo {pages} {075204}
  (\bibinfo {year} {2010})}\BibitemShut {NoStop}%
\bibitem [{\citenamefont {Ghimire}\ \emph {et~al.}(2010)\citenamefont
  {Ghimire}, \citenamefont {DiChiara}, \citenamefont {Sistrunk}, \citenamefont
  {Agostini}, \citenamefont {DiMauro},\ and\ \citenamefont
  {Reis}}]{Ghimire2010}%
  \BibitemOpen
  \bibfield  {author} {\bibinfo {author} {\bibfnamefont {S.}~\bibnamefont
  {Ghimire}}, \bibinfo {author} {\bibfnamefont {A.~D.}\ \bibnamefont
  {DiChiara}}, \bibinfo {author} {\bibfnamefont {E.}~\bibnamefont {Sistrunk}},
  \bibinfo {author} {\bibfnamefont {P.}~\bibnamefont {Agostini}}, \bibinfo
  {author} {\bibfnamefont {L.~F.}\ \bibnamefont {DiMauro}}, \ and\ \bibinfo
  {author} {\bibfnamefont {D.~A.}\ \bibnamefont {Reis}},\ }\href {\doibase
  10.1038/nphys1847} {\bibfield  {journal} {\bibinfo  {journal} {Nat. Phys.}\
  }\textbf {\bibinfo {volume} {7}},\ \bibinfo {pages} {138} (\bibinfo {year}
  {2010})}\BibitemShut {NoStop}%
\bibitem [{\citenamefont {Keldysh}(1958)}]{Keldysh1958}%
  \BibitemOpen
  \bibfield  {author} {\bibinfo {author} {\bibfnamefont {L.~V.}\ \bibnamefont
  {Keldysh}},\ }\href@noop {} {\bibfield  {journal} {\bibinfo  {journal} {Sov.
  Phys. JETP}\ }\textbf {\bibinfo {volume} {33}},\ \bibinfo {pages} {763}
  (\bibinfo {year} {1958})}\BibitemShut {NoStop}%
\bibitem [{\citenamefont {Franz}(1958)}]{Franz1958}%
  \BibitemOpen
  \bibfield  {author} {\bibinfo {author} {\bibfnamefont {W.}~\bibnamefont
  {Franz}},\ }\href@noop {} {\bibfield  {journal} {\bibinfo  {journal} {Z.
  Naturforsch. A}\ }\textbf {\bibinfo {volume} {13}},\ \bibinfo {pages} {484}
  (\bibinfo {year} {1958})}\BibitemShut {NoStop}%
\end{thebibliography}%
\end{document}